\documentclass[twocolumn,prb,showkeys,graphicx,superscriptaddress]{revtex4}
\usepackage{graphicx}
\usepackage{dcolumn}
\usepackage{bm}
\usepackage[cmex10]{amsmath}
\usepackage[utf8]{inputenc}
\usepackage{float}
\usepackage{amssymb}
\usepackage[T1]{fontenc}
\usepackage{natbib,hyperref}
\usepackage{color,soul}

\usepackage[utf8]{inputenc}
\usepackage{textcomp}
\usepackage{palatino}

\usepackage{graphicx}
\usepackage{epsfig}

\begin{document}

\preprint{APS/123-QED}

\title{Electric-field tunable spin waves in PMN-PT/NiFe heterostructure: experiment and micromagnetic simulations}

\author{Slawomir Zi\k{e}tek}
\email{zietek@agh.edu.pl}
\affiliation{AGH University of Science and Technology, Department of Electronics, Al. Mickiewicza 30, 30-059 Krak\'{o}w, Poland}

\author{Jakub Ch\k{e}cinski}

\affiliation{AGH University of Science and Technology, Department of Electronics, Al. Mickiewicza 30, 30-059 Krak\'{o}w, Poland}
\affiliation{AGH University of Science and Technology, Faculty of Physics and Applied Computer Science, Al. Mickiewicza 30, 30-059 Krak\'{o}w, Poland}

\author{Marek Frankowski}

\affiliation{AGH University of Science and Technology, Department of Electronics, Al. Mickiewicza 30, 30-059 Krak\'{o}w, Poland}

\author{Witold Skowronski}

\affiliation{AGH University of Science and Technology, Department of Electronics, Al. Mickiewicza 30, 30-059 Krak\'{o}w, Poland}

\author{Tomasz Stobiecki}
\affiliation{AGH University of Science and Technology, Department of Electronics, Al. Mickiewicza 30, 30-059 Krak\'{o}w, Poland}

\date{\today}

\begin {abstract}

We present a comprehensive theoretical and experimental study of voltage-controlled standing spin waves resonance (SSWR) in PMN-PT/NiFe multiferroic heterostructures patterned into microstrips. A spin-diode technique was used to observe ferromagnetic resonance (FMR) mode and SSWR in NiFe strip mechanically coupled with a piezoelectric substrate. Application of an electric field to a PMN-PT creates a strain in permalloy and thus shifts the FMR and SSWR fields due to the magnetostriction effect. The experimental results are compared with micromagnetic simulations and a good agreement between them is found for dynamics of FMR and SSWR with and without electric field. Moreover, micromagnetic simulations enable us to 
discuss the amplitude and phase spatial distributions of FMR and SSWR modes, which are not directly observable by means of spin diode detection technique.

\end{abstract}

\maketitle

\section{Introduction}

Information processing, transmission and storage in modern electronic devices have been related to the electric charge flow and its accumulation. However, utilizing spin waves (SW) propagating through the ferromagnetic medium as information carriers can lead to significant reduction of current-flow-related energy losses, thereby improving efficiency and speed of binary operations\cite{jamali2013spin}.
SW can be excited in ferromagnetic material cuased by, e.g. radio frequency (RF) magnetic field \cite{yu2014magnetic,jamali2013spin,yu2013omnidirectional}, electric-field-induced magnetoelastic anisotropy changes \cite{cherepov2014electric}, as well as spin torque\cite{demidov2010direct}. On the other hand, electric-field-modified magnetization dynamics (e.g. via voltage control effective anisotropy) allows for electrical control of SW propagation in ferromagnet\cite{cherepov2014electric}. Thereby, an electrical detection and generation of spin waves have become an attractive alternative to charge-based electronics from an application point of view. 
Newly designed spintronics devices based on SW such as SW logic gates  \cite{jamali2013spin,chen2015control}, multiferroic SW generators\cite{cherepov2014electric}, SW multiplexers\cite{vogt2014realization}, couplers\cite{yu2013omnidirectional} and waveguides\cite{yu2014magnetic} have been currently strongly developed. A particular case of SW are  standing spin waves (SSW), produced by an interference between waves reflected back and forth. Standing spin waves resonance (SSWR) have been intensively investigated by Brillouin light scattering \cite{urazhdin2014nanomagnonic} and also by local ferromagnetic resonance (FMR) excitation\cite{bai2011spin,banholzer2011visualization,yu2013omnidirectional,schoeppner2014angular}. Yamaguchi et al.\cite{yamaguchi2007rectification} as a first have shown a spin diode (SD) detection of FMR/SSWR in NiFe microstrips on a silicon substrate. Permalloy microstrips provide an excellent ferromagnetic medium to study a nature of SSRW because of relatively low Gilbert damping parameter ($\alpha$ $\approx$ 0.007) and high anisotropic magnetoresistance (AMR) ratio (up to 4$\%$)\cite{mcguire1975anisotropic,stobiecki1975preparation}. These two important features are required for excitation and detection of magnetization dynamics by means of spin diode effect\cite{tulapurkar2005spin,zietek2015rectification}. In this work, we demonstrate a voltage control of SSWR and also FMR in NiFe microstrip deposited on PMN-PT piezoelectric substrate. SSWR are excited with an alternatig magnetic field produced by RF currnet, whereas a purely electrical detection is based on SD effect. Voltage applied to the piezoelectric substrate creates stress in ferromagnetic film, which due to the inverse magnetostriction effect induces the magnetoelastic anisotropy \cite{kim2013coherent}, and consequently changes FMR and SSWR fields.

\section{Experimental}

On a polished PMN-PT (0.72Pb(Mg$_{1/3}$Nb$_{2/3}$)- O$_3$–0.28PbTiO$_3$) [011] oriented piezoelectric substrate, a 20 nm of Ni$_{80}$Fe$_{20}$  was deposited using magnetron sputtering. In order to apply electric field perpendiculary to the piezoelectric substrate, the bottom side of the crystal was covered by a Ti 5 nm/Au 50 nm layer. Microstrips of 6.7 \textmu m width and 90 \textmu m length were fabricated using electron beam lithography and ion-beam etching. Angular depenence of SD-FMR apmlitudes was measured in a four-pole electromagnet probe station, which enables application of an arbitrary field vector in the sample plane. FMR spectra were measured using (SD) technique similarly to Ref. \onlinecite{zietek2015rectification}. An RF current of 10 dBm power at frequencies in range form 3 GHz to 8 GHz was passed trough the ferromagnetic microstrip. Magnetization oscillations induced by this current, due to the AMR effect, cause small changes of resistance in microwave frequency range, which mixes with RF current and givea a raise of SD voltage. This voltage was measured using a lock-in technique during the magnetic field sweep at various magnetic field angles $\theta_H$. A top and bottom layer of PMN-PT piezoelectric substrate were previously covered by 50nm of Au, excluding inconsiderable area where micro-devices were fabrication. In order to generate in-plane stress in piezoelectric substrate, a 100 V was applied from additional voltage source to top and bottom metallic layer, simillary as in Ref. \onlinecite{2016zietekapl}.

\section{Results and discussion}

 \subsection{Angular dependence of FMR and SSWR}
Fig.1a. shows SD-FMR/SSWR spectra obtained for 0$^{\circ}$ $\leq$ $\theta_H$ $\leq$ 90$^{\circ}$ magnetic field angles.

\begin{figure}[h!]
\begin{center}
	\includegraphics[width=9cm]{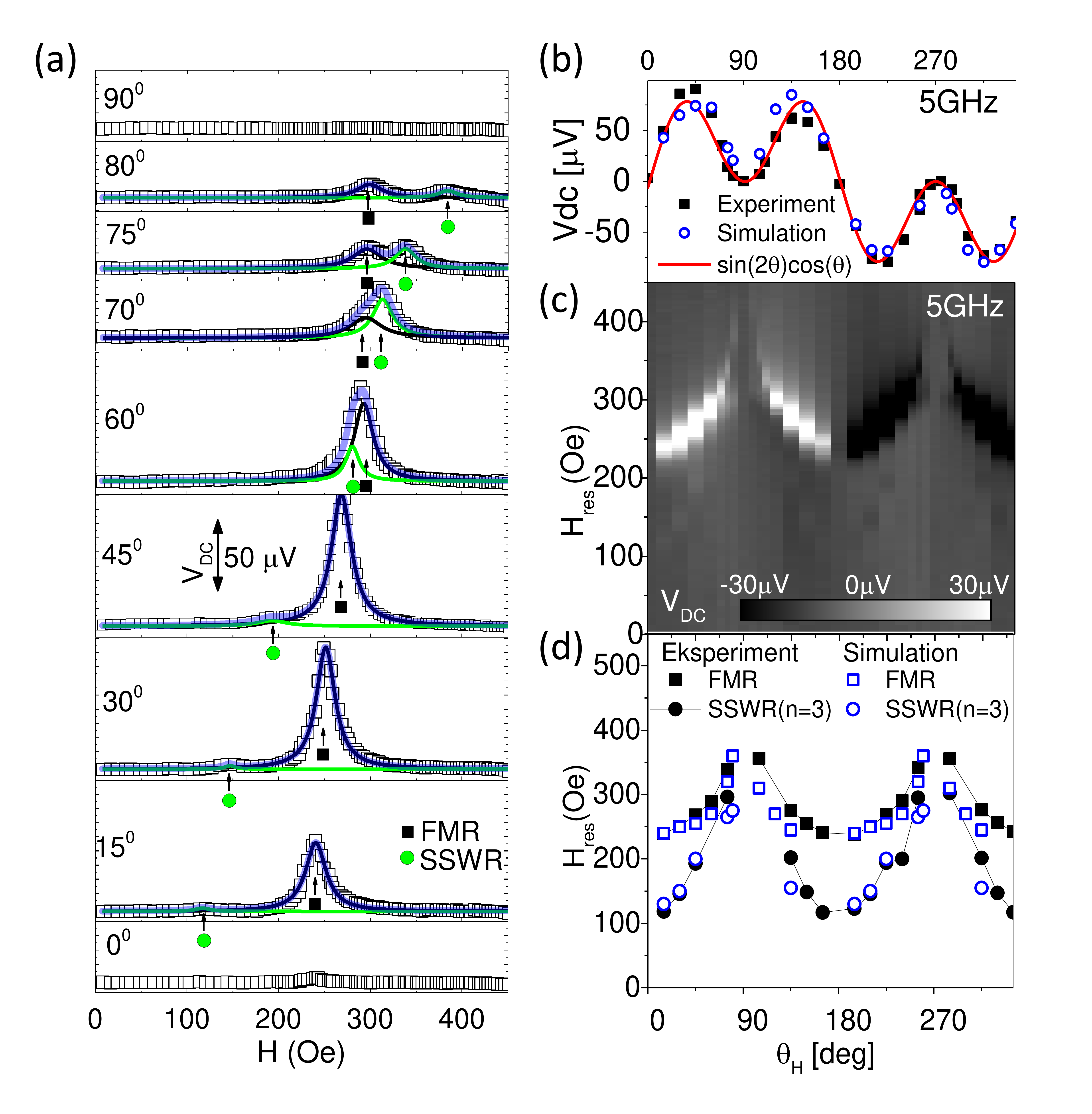}
	\caption{ (a) FMR and SSWR spectra measured at constant frequency (f=5GHz) as a functions of magnetic field at chosen angles $\theta_H$ (arrows indicate SSWR). Black and green solid lines represent Lorentzian function fits to FMR and SSWR, respectively. Semi-transparent blue line represents sum of FMR and SSWR fits, (b) angular dependence of SD-FMR amplitudes, (c) 2D-plot of SD voltage as a function of magnetic field magnitude and angle $\theta_H$. The grey scale represents values of SD voltage. (d) Angular dependences of FMR and SSWR resonance fields.}
\label{fig:RH}
	\end{center}
\end{figure}

Black and green solid lines represents Lorentzian function fit to the FMR and SSWR, respectively, and semi-transparent blue line represents a sum of both fit lines. 
Measured and simulated SD-FMR amplitudes as a function of $\theta_H$ are compared in Fig.1b and both of them well satisfy the relation $sin(2\theta_H)cos(\theta_H)$ proposed in Ref.\onlinecite{yamaguchi2007rectification}. Fig.1c. shows a 2D-plot of SD voltage as a function of magnetic field angle and magnitude. Apart from the uniform FMR mode, in this plot, also the SSWR can be seen as a trace marked below FMR.  The resonance fields of FMR and SSWR as a function of $\theta_H$ are shown in Fig.1d. The $\theta_H=40 ^{\circ}$ was selected for investigation of an electric field control of SSWR and FMR resonance fields, due to good separation between peaks and high SD effect efficiency.

\subsection{Electric-field control of SSWR}

\begin{figure}[h!]
\begin{center}
	\includegraphics[width=9cm]{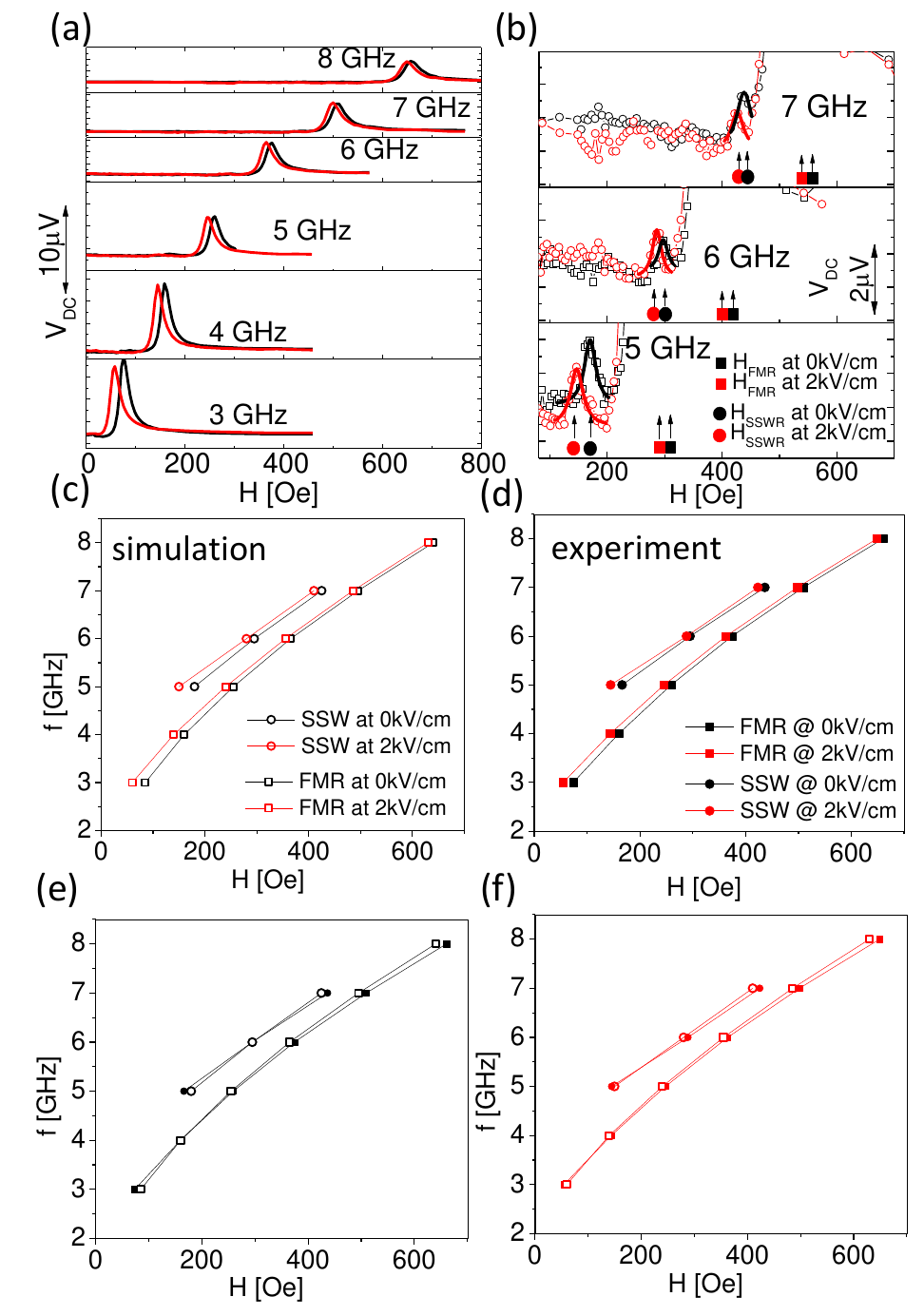}
	\caption{ (a) FMR (b) SSWR spectra measured without (black lines) and with (red lines) electric field of 2 kV/cm. (c) Simulation and (d) experimental Kittel relations with and without electric field. Comparison of simulation (empty) and experimental (filled) Kittel relations without (e) and with (f) electric field.}
\label{fig:RH}
	\end{center}
\end{figure} 

Examples of the electric-field tunable FMR spectra are shown in Fig.2a. Typically, the amplitude of the SD voltage in the region of SSWR is about one order of magnitude lower than SD-FMR voltage amplitude. Fig.2b. shows magnified SSWR spectra.  Dispersion relations of FMR and SSWR are shown in Fig.2c. Black line represents measurements without electric field, while red line represents measurements for the electric-field of 2 kV/cm. The shift of the FMR for the applied electric-field is the most visible at lower magnetic fields and frequencies because of relatively low Zeeman magnetic energy contribution to the total energy term \cite{2016zietekapl}. The most significant effect of electric field on SSWR was obtained for 5 GHz. For higher frequencies/fields the shifts of SSWR were smaller due to higher contribution of Zeeman energy which compete with additional magnetoelastic energy. Despite the low magnetostriction of $Ni_{80}Fe_{20}$ ($\lambda\approx2.5\times10^{-6}$) \cite{hill2013whole}, an application of electric field (2 kV/cm) generates relatively strong SSWR shift of 22 Oe.

Micromagnetic simulation, described in the next paragraph, truly reflects dispersion relations shift (Fig.2d). Fig.2e and Fig.2f quantitatively compare experimental and simulation data obtained for 0 and 2 kV/cm, respectively. Experimental and simulation spectra for constant frequency of 5 GHz and electric field of 0 and 2kV/cm are compared in Fig.3a and Fig.3b, respectively.

\begin{figure}[h!]
\begin{center}
	\includegraphics[width=7cm]{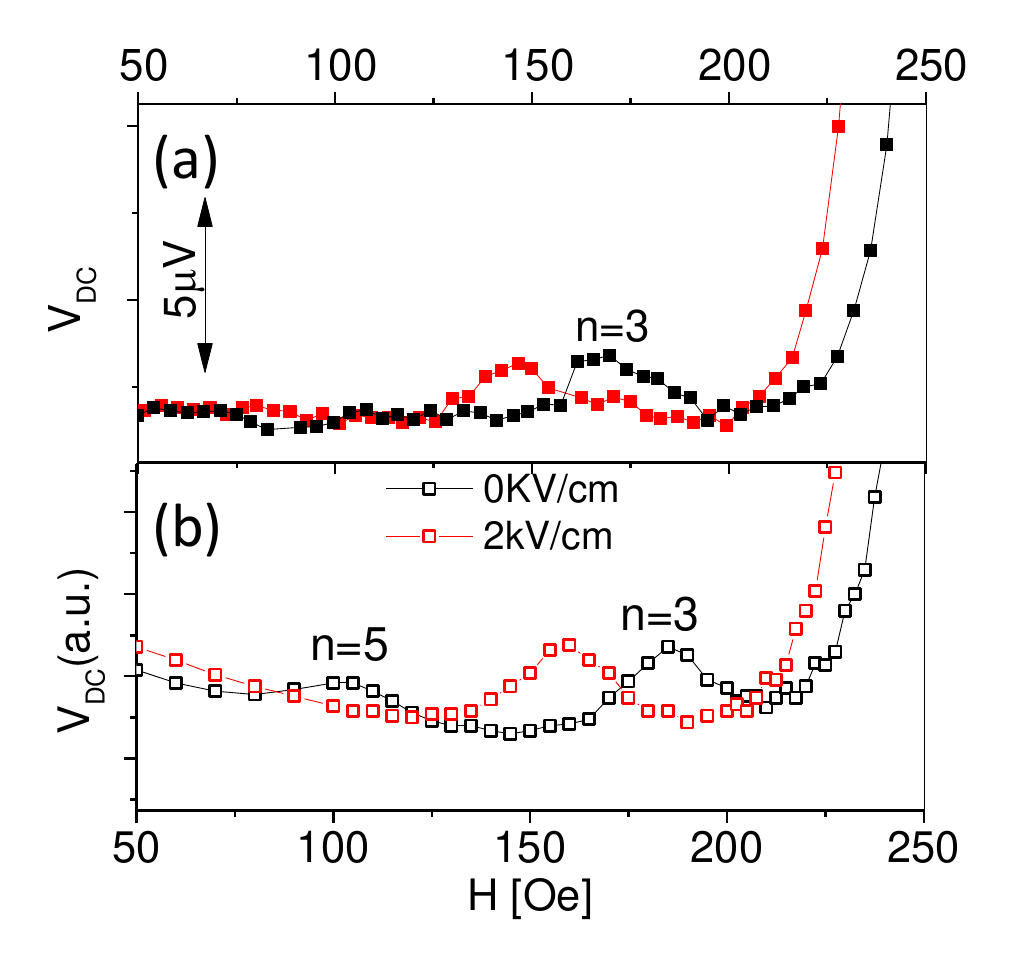}
	\caption{(a) Experimental and (b) simulation FMR and SSWR spectra measured at $\theta_H=40^\circ$ at constant frequency of 5 GHz at 0 and 2kV/cm of electric field. }
\label{fig:RH}
	\end{center}
\end{figure}

\section{Micromagnetic simulations}

In order to confirm that the measured modes are SSWRs and identify thier harmonic number, we performed a set of micromagnetic simulations of FMR and SSWR angular dependences and their behaviour in an electric field. We utilized OOMMF software \cite{donahue1999oommf} together with our custom extensions \cite {frankowski2015spatial,chkecinski2016mage} allowing for fast and efficient preparation of the  simulation files and analysis of the obtained results. We assumed the following parameters for the NiFe strips: ferromagnetic exchange constant equal to 1.3 $\cdot$ $\mathrm{10^{-11}}$ J/m, damping constant $\alpha$ = 0.007, mesh qubic cell size equal to 5 nm. Saturation magnetization $M_S$ was given so that $\mu_0 M_S$ = 1.2 T. A small uniaxial anisotropy of $K_P$ = 785 J/$\mathrm{m^3}$ along the hard in-plane axis was used. The shape anisotropy is taken into account by means of demagnetization field calculations during simulation process. The LLG equation\cite{landau1935theory,gilbert2004phenomenological} was solved numerically using the Runge–Kutta method. To be able to use OOMMF for modelling of our relatively large device, we simulated strips with full 6.7 \textmu m width (y axis) and 20 nm thickness (z axis) and used finite periodic boundary conditions in the longest (90 $\mathrm{\mu m}$) dimension. It was performed by calculating the modified demagnetization tensor as introduced in the OOMMF extension described in Ref. \onlinecite{wang2010two}. Because of that, we considered only a 100 nm long (x axis) part of the strip, which was nonetheless sufficient for the purpose of SSWR analysis since they did not propagate along the longest x axis.

In the FMR and SSWR simulations, we dynamically calculated the total AMR ratio for a strip based on the magnetization vectors distribution in each time step. It was done using an approach similar to the one presented in Ref. \onlinecite{frankowski2014micromagnetic}, namely we treated cells along the x dimension as connected in series and then the resultant channels as connected in parallel. The total resistance was then mixed with the current in order to produce the SD voltage output. By sweeping the magnetic bias field values for different angles, we were able to localise FMR and SSWR for an example frequency of 5 GHz (see Fig. 1d) and to reproduce the $\mathrm{V_{dc}}$ dependence on the angle for a given bias field value (see Fig. 1b). 

During the electric field simulations, we modeled the effects of the external electric field as uniaxial anisotropy changes assuming that a 1 kV/cm electric field decreases the anisotropy along the in-plane hard axis by 635 J/$\mathrm{m^3}$ and introduces a small additional anisotropy of 80 J/$\mathrm{m^3}$ along the in-plane easy axis.

 The inclusion of this effect caused both FMR and SSWR values for a given frequency to move towards smaller magnetic bias fields, which agrees with the experimental results (see Fig. 2). An example of a simulated SSWR line and its shift induced by the electric field is depicted in Fig. 3b.

\begin{figure}[h!]
\begin{center}

	\includegraphics[width=9.3cm]{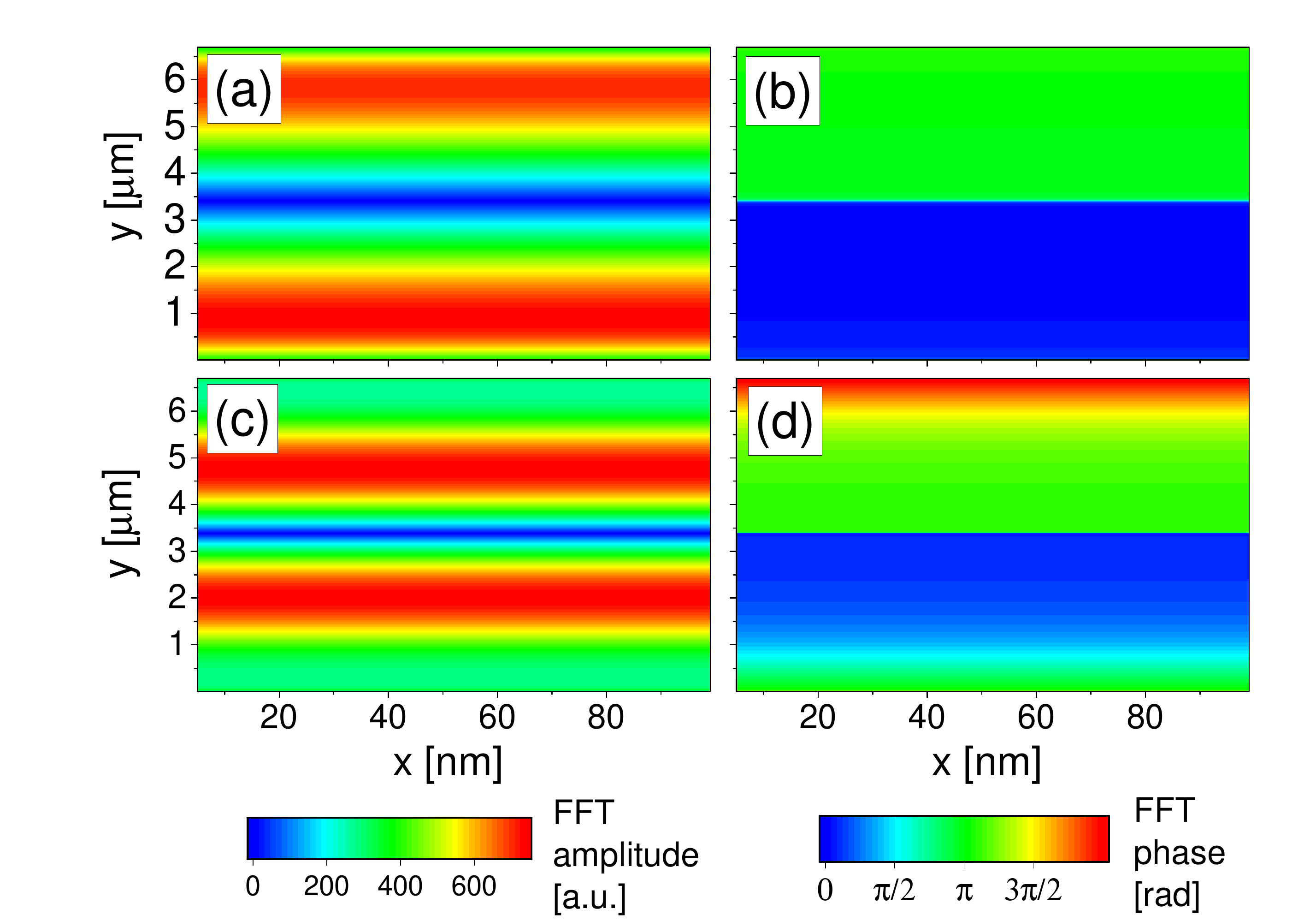}
	\caption{Spatial distribution maps of (a) FFT amplitude of FMR mode, (b) FFT phase of FMR mode, (c) FFT amplitude of SSWR 3rd mode, (d) FFT phase of SSWR 3rd mode. }
\label{fig:RH}
	\end{center}
\end{figure} 

In all simulations described so far, the magnetization was excited by an Oersted field. We assumed the existence of the field produced by an electric current flowing through a strip under some sort of symmetry-breaking conditions, for example difference in Sondheimer Fuchs electron scattering at top and bottom NiFe interfaces \cite{thiaville2008electrical}. We note that if the current distribution is perfectly symmetric, there are no uncompensated Oersted field present in the system and rectification of a non-zero output signal is impossible \cite{zietek2015rectification,thiaville2008electrical}. On the other hand, if one allows the current distribution to produce an uncompensated field, the SD signal will arise. We used a simple assumption where the amplitude of RF current flowing through the upper half of the stripe (from 10 nm to 20 nm on the z axis) was twice as large as the current flowing through the lower half (from 0 nm to 10 nm). Although this approach does not allow for an easy quantitative comparison of the $\mathrm{V_{dc}}$ signal amplitude with the experimental results, it should be sufficient to reproduce qualitative features of the FMR and SSWR lines. Fig.4. presents spatial distribution maps of both the FMR and the SSWR mode under their respective magnetic bias fields at constant frequency of 5 GHz, obtained using technique and software described in Ref. \onlinecite{frankowski2015spatial}. One can see, however, that the suspected FMR mode (Fig.4a for amplitude and 4b for phase) does not display the features one would expect from a genuine ferromagnetic resonance, but instead indicates a more complicated dynamics. We attribute this behavior to the character of the nature of the Oersted field excitation, which, even without the symmetry-breaking present, introduces a constant phase difference of $\mathrm{\pi}$ between two halves of the strip (as seen in Fig.4b). Therefore, although the simulations we performed so far corresponded to the actual experimental procedure, they did not provide us with information sufficient to identify the exact character of FMR and SSWR modes. 

\begin{figure}[h!]
\begin{center}
	\includegraphics[width=9cm]{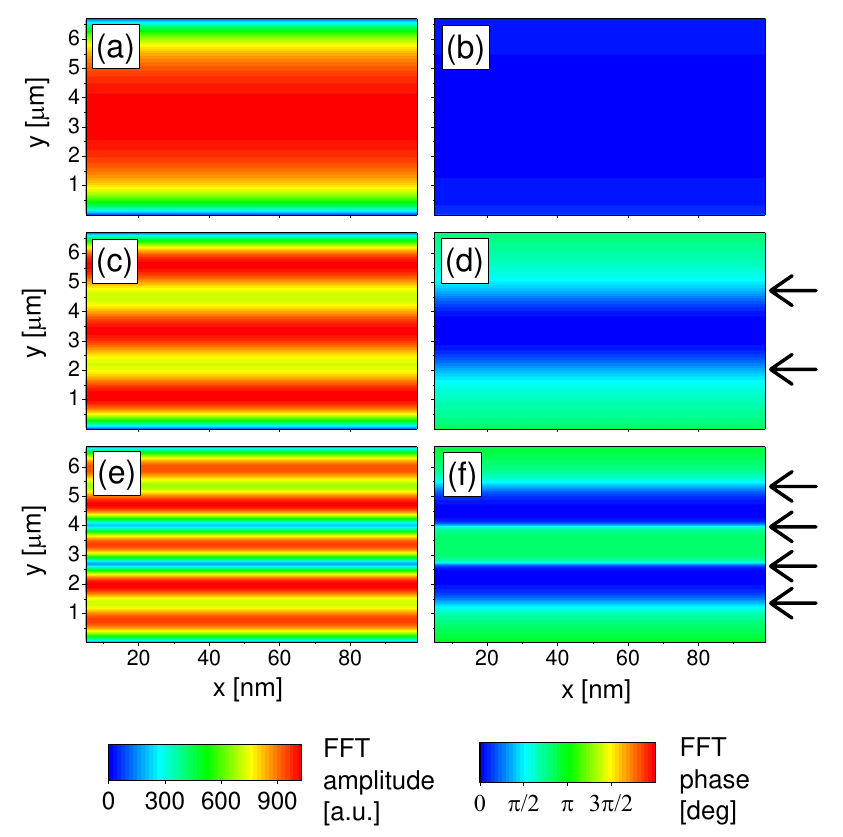}
	\caption{Spatial distribution maps of (a) FFT amplitude of FMR mode, (b) FFT phase of FMR mode, (c) FFT amplitude of the 3rd SSWR mode, (d) FFT phase of the 3rd SSWR mode, (e) FFT amplitude of the 5th SSWR mode, (f) FFT phase of the 5th SSWR mode. Black arrows indicates SSWR nodes.}
\label{fig:RH}
	\end{center}
\end{figure} 
 
To confirm this hypothesis, we performed an additional set of simulations with the uncompensated magnetic field introduced artificially as a spatially uniform field. While not corresponding to the experimental procedure, this assumption eliminated the problem of Oersted-originating phase difference and allowed us to further investigate the character of our resonance modes. We found that the resonance fields at 5 GHz frequency of the suspected FMR and SSWR modes remained unchanged with the new kind of excitation, but their spatial distribution changed, as presented in Fig.5. One can see that the first mode (Fig.5a and 5b) presents now a clear ferromagnetic resonance character with a nearly uniform amplitude and virtually no phase difference across the whole sample. The second mode (Fig.5c and 5d) has also a more clear character when compared to the results from Fig.4 and can be identified as a 3rd harmonic standing spin wave. Additionally, at a lower bias field value we were able to identify a smaller, 5th harmonic mode (Fig.5e and 5f), which was not clearly visible in the experimental technique. We conclude that the results obtained from uniform field simulations indicate the actual physical nature of the investigated modes, which was partially obscured when the experimental-like, Oersted field excitation was used. Therefore, the first and the second mode observed in the experimental part of this work can indeed be described as the FMR and the SSWR mode, respectively.

\section{Summary}

Magnetization dynamics in NiFe/PMN-PT multiferroics heterostructures was investigated using the SD effect. Firstly, the SD-FMR/SSWR spectra were measured at different field angles, and in the next step, voltage control of FMR and SSWR resonance fields was investigated. An electric field was applied across the PMN-PT substrate to create a piezoelectric strain, which due to the magnetostriction effect influenced the magnetoelastic anisotropy of the NiFe film. Changes of the magnetic energy resulted in a shift of both FMR and SSWR resonance fields. Experimental results were compared with micromagnetic simulations, and a good quantitative agreement between them was found. Spatial distribution maps of the magnetization response to both an Oersted-like excitation and a uniform excitation were produced and the FMR/SSWR character of the induced modes was analysed. The harmonic number of the investigated SSWR mode was also identified.

Further research on the magnetization dynamics in multiferroics heterostructures will be necessary for development of a new type of voltage-tunable microwave filters or detectors with a very broad frequency range.

\section*{Acknowledgement}
We acknowledge Polish National Science Center grant Harmonia-UMO-2012/04/M/ST7/00799. S.Z. acknowledges Dean's grant no. 15.11.230.277, W.S. acknowledges Polish National Center for Research and Development grant No. LIDER/467/L-6/14/NCBR/2015, J.Ch. acknowledges the scholarship under Marian Smoluchowski Krakow Research Consortium KNOW programme. Numerical calculations were supported by PL-GRID infrastructure.

\bibliographystyle{nature}


\bibliography{bibliography}

\end{document}